\documentclass[twocolumn,showpacs,preprintnumbers,amsmath,amssymb,prl]{revtex4}
\usepackage[pdftex]{graphicx}% Include figure files
\begin{document}

\title{Wide defect in a Resonantly Absorbing Bragg Grating
as a nonlinear microresonator for polaritons }
\author{Elena V. Kazantseva }
\email{elena@math.arizona.edu}\affiliation{Department of
Mathematics, University of Arizona, 617 N. Santa Rita Ave., Tucson,
AZ 85721, USA}
\author{Andrei I. Maimistov}
\email{maimistov@pico.mephi.ru}\affiliation{Department of Solid State Physics, Moscow Engineering
Physics Institute, Russia}

\date{\today}

\begin{abstract}
The nonlinear polariton transmission, reflection and trapping by a
defect in the resonantly absorbing Bragg grating (RABG) is
demonstrated in numerical simulation. It is shown that the wide
defect under some conditions could effectively serve as
microresonator for polaritonic wave storage. The three types of the
defect such as microcavity, groove and stripe are considered.
Capture the electromagnetic field inside the microcavity (with no
resonant nanoparticles) placed in the RABG is observed, as well as
stuck of trapped polarization modes to the defect edges for the
groove (defect span with reduced density of nanoparticles) and for
the stripe with relatively increased density. Strong radiation
reflection and adhered propagation of the polarization mode along
the first edge of the stripe with high density of resonant atoms is
exhibited by numerical computation.
\end{abstract}


\maketitle

%\textbf{Introduction.}
In last years the interest to the intermediate between electronics
and photonics has arisen. The new terms polaritonics and plasmonics
attract attention of scientists and engineers. The interest of
polaritonics is in interaction of light with the microstructured
matter. Due to the prominent progress that has occurred last years
in fabrication of novel materials with engineered internal structure
(photonic bandgap fibers, carbon nanotubes, nanowires), there is the
demand for an analysis of the phenomena that can arise due to
specific properties of artificial materials.

The substantial achievements in fabrication of periodic structures
in different materials attract interest to metallic periodic
structures such as metallic combs, periodic array of
indentations~\cite{1}, two-dimensional hole lattices~\cite{2} and
one-dimensional groove arrays~\cite{3} machined into flat
interfaces, conducting wire tailored with a periodic array of radial
grooves ~\cite{4} and even 'superstructures of nanowires and
nanoparticles connected by molecular springs'~\cite{5}. The objects
of interest are waves resulting from the photon interaction with
plasmonic oscillations in metal ~\cite{7}. The theory of surface
plasmons polaritions on structured surfaces as well as recent
advances in experimental studies are reported ~\cite{8}. The
relevant method for preparation the dielectric structure with
periodic thin films containing metallic nanoparticles or molecules
could be e-beam deposition or layer-by-layer adsorption technique
~\cite{9} which allows creation of nanometer-scale multilayered
films.

The nonlinear periodic structure that is referred to as resonantly
absorbing Bragg grating (RABG) has been proposed and investigated
in~\cite{10,12,16}. The RABG consists of linear homogeneous
dielectric medium containing thin film array of resonant two-level
atoms. The results of the theory are summarized in ~\cite{17}. The
model developed in~\cite{18,19,19r} describes the ultrashort
(comparing to the relaxation time) pulse which is propagates through
the regular Bragg grating evolved from films of metallic
nanoparticles. Nonlinear plasmonic oscillations are governed by the
cubic Duffing equation. Solitary electromagnetic waves coupled with
media polarization, i.e., polaritonic solitary waves, were found.
The robustness of these \emph{polaritonic gap solitons} (PGSs) was
demonstrated by numerical simulation.

Optical soliton scattering on defects in lattices is attractive
problem of nonlinear optics. The paper~\cite{23} is devoted to
trapping of light in a nonuniform resonant structure by the defect
which is created as a result of the inversion of the atoms
population. Depending on strength of the defect the pulse either can
pass through it with low radiation or to localize in the defect.
Action of the second pulse could lead to depinning of the initial
pulse or to trapping both of them in the defect. Trapping Bragg
solitons by a pair of localized defects has been also
demonstrated~\cite{24}. Thus, it would be very natural to consider
interaction of the PGSs with different types of the RABG defects,
such as the absence of the resonant atom layers in the grating or
oppositely the conglutination of several films. The influence of all
these effects arising in the process of artificial media fabrication
can be crucial in material guiding properties.

A homogeneous span between two distributed Bragg mirrors is the
effective laser microresonator in integrated optics~\cite{25,26} and
nano-optics~\cite{27,28,29}. During last years the persistent
interest for optical properties of these structures is subsists. The
review of the nonlinear properties of semiconductor microcavities
can be found in~\cite{30}. The quantum well that is grown inside the
Bragg-mirror microcavity appears as the defect in regular periodic
structure. So, the array of quantum wells in Bragg-mirror
microcavity is tantamount to wide defect in regular grating. A
similar statement is true also for the resonantly absorbing Bragg
grating.

In this paper we will consider the three types of microresonator,
which will be referred as microcavity, groove and stripe.
Specification of these terms will be done below. We perform
numerical simulation of the coupled solitary waves which are
interact resonantly with media consisting of dielectric layers
alternating with thin films with metallic nanoparticles (or quantum
dots, nanoagregates with nonlinear dielectric properties)
~\cite{18}. Due to Bragg resonance condition a weak electromagnetic
pulse can not propagate in the RABG, whereas the moving polaritonic
gap soliton that has been found in ~\cite{18,19,19r} is able to
reach for inner defect.

%%\textbf{Model description.}
The model of the polaritonic wave propagating in the resonantly
absorbing Bragg grating is considered following to~\cite{18}. The
width of the resonant film is about hundred of nanometers and it is
negligible small comparing to the dielectric layer separating two
thin films of resonant inclusions. According to slowly varying
envelope approximation,pulse width includes about hundred of grating
double periods and varies slowly on one period. Considering
microresonator as the defect in the RABG, we assume that this defect
consists of several hundreds of layers containing resonant
nanoparticles so the slowly varying envelope approximation is still
valid.  The normalized form of the model equations is
\begin{eqnarray}
i\left( \frac{\partial }{\partial \zeta }+\frac{\partial }{\partial \tau }%
\right) e_{1}+\delta e_{1}=-\gamma \left( \zeta \right)p  \notag \\
i\left( \frac{\partial }{\partial \zeta }-\frac{\partial }{\partial \tau }%
\right) e_{2}-\delta e_{2}=+\gamma \left( \zeta \right)p  \notag \\
i\frac{\partial }{\partial \tau }p+\Delta p+\mu \left\vert
p\right\vert ^{2}p=-\left( e_{1}+e_{2}\right).\label{eq1}
\end{eqnarray}
The dimensionless variables  $e_{1,2}$ are slowly varying envelopes
of the electromagnetic field components, which are propagate in
forward and backward directions, and $p$ is the slowly varying
envelope of the polarization, which is determined by the array of
thin films containing resonant nanoparticles. Here $\Delta
=2\sqrt{\varepsilon}( \omega _{d}-\omega _{0}) /\omega _{p}$ is the
detuning of a nanoparticle's resonance frequency from the field's
carrier frequency  $\omega _{0}$, $\omega _{p}$ is the plasma
frequency, $\omega _{d}$ is the frequency of dimensional
quantization of nanoparticles, $\delta =2( c/\omega _{p}) \Delta
q_{0}$ with $\Delta q_{0}=q_{0}-2\pi /a$, $q_{0}=\omega
_{0}\sqrt{\varepsilon }/c$ is wave vector in the media, and
$\varepsilon $ is permittivity, $\mu $ is a dimensionless
coefficient of anharmonicity. Normalized spatial and time variables
denote as $\zeta $ and  $\tau $, respectively. Total field acting on
resonant atoms is the sum of  $e_1$ and $e_2$ ~\cite{18}.

\begin{figure}[b]
%\resizebox {width=8.5cm}{length=4.5cm}
 \centering\includegraphics[width=8.5cm]{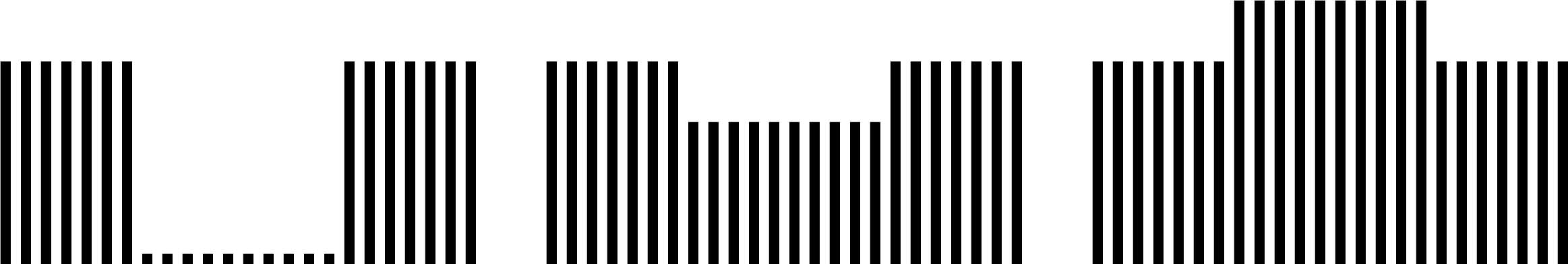}
\caption{{\protect\footnotesize {Distribution of nanoparticles
density in (a) RABG microcavity  $\gamma (\zeta )=0 $, (b) RABG wide
groove defect $\gamma (\zeta )<1$ and (c) RABG wide stripe defect
$\gamma (\zeta )>1.$}}} \label{Fig1}
\end{figure}

The defects in RABG are described by the indicator function $\gamma(
\zeta)$. The regular RABG is produced from thin layers with equal
concentration of nanoparticles. For this reason we put $\gamma (
\zeta)=1 $ in the regular grating. Here we consider the lattice
defect as span with alternative nanoparticles concentration (see
Fig.~\ref{Fig1}). When coordinate $\zeta $ belongs to defect region
the function $\gamma ( \zeta ) $ is not equal unit. If the density
of nanoparticles in layer at point $\zeta $ is more (less) comparing
with the density of particles in regular part of the grating, then
$\gamma(\zeta)>1 $($\gamma(\zeta)<1$). In other words, the indicator
function is equal to ratio of nanoparticle density in the defect
layer per nanoparticle density of regular layers. The region with
$\gamma(\zeta)<1 $ will be termed the groove defect. The region
where ($\gamma(\zeta)>1$) will be denoted as the stripe.
The solitary wave solution (i.e.,the polaritonic gap soliton)
corresponding to regular RABG has been found and discussed
in~\cite{18,19,19r}. It reads as
\begin{eqnarray}
e_{1}\left( \eta \right) =-0.5\left( 1+\alpha \right) f_{s}\left(
\eta \right) \exp \left\{ i\delta \tau \right\} ,\notag \\
e_{2}\left( \eta \right) =-0.5\left( 1-\alpha \right) f_{s}\left(
\eta \right) \exp \left\{ i\delta \tau \right\} ,\notag \\
p\left( \eta \right) =q\left( \eta \right) \exp \left\{ i\delta \tau
\right\}. \label{eq21}
\end{eqnarray}
where  $\eta =\tau -\alpha \zeta $,  $\beta =2/( \alpha ^{2}-1) $,
$\alpha >1$ is a free  parameter defining the solitary wave group
velocity. Auxiliary functions   $f_{s}=u\exp \{ i\varphi \}$ and
$q=r\exp \{ i\psi \}$ are defined through the following amplitudes
\begin{eqnarray}
u^{2}\left( \eta \right) ={4 \sqrt{\beta }\beta}{\cosh^{-1} \left[ 2\sqrt{%
\beta }\left( \eta -\eta _{0}\right) \right] }, \notag \\
r^{2}\left( \eta \right) ={4\beta }{\cosh^{-1} \left[ 2\sqrt{\beta
}\left( \eta -\eta _{0}\right) \right] }. \label{eq3}
\end{eqnarray}
and phases
\begin{eqnarray}
\varphi \left( \eta \right) =\varphi _{0}\pm \arctan \tanh \left[ \sqrt{%
\beta }\left( \eta -\eta _{0}\right) \right] , \notag \\
\psi \left( \eta \right) =\psi
_{0}\pm 3\arctan \tanh \left[ \sqrt{\beta }\left( \eta -\eta _{0}\right) %
\right]. \label{eq4}
\end{eqnarray}

Initial phases are set in a such way that $\varphi _{0}-\psi_{0}=\pi
/2$, at  $\eta \rightarrow -\infty $. In (\ref{eq3}) and (\ref{eq4})
parameter  $\eta_0 $ is an integration constant that indicates the
initial pulse position.

The nonlinear polaritonic wave (\ref{eq21}) propagates through the
ideal RABG without radiation loss or change of the initial shape. It
should be noted that passing the RABG is not allowed for linear
waves due to the band gap. For this reason we used the solution
(\ref{eq21}) as the initial configuration for our numerical
simulation of the solitary wave propagation through the
microresonator span in RABG. (Fig.~\ref{Fig1}).
\begin{figure}[b]
%\resizebox {width=8.5cm}{length=4.5cm}
\centering\includegraphics[width=8.5cm]{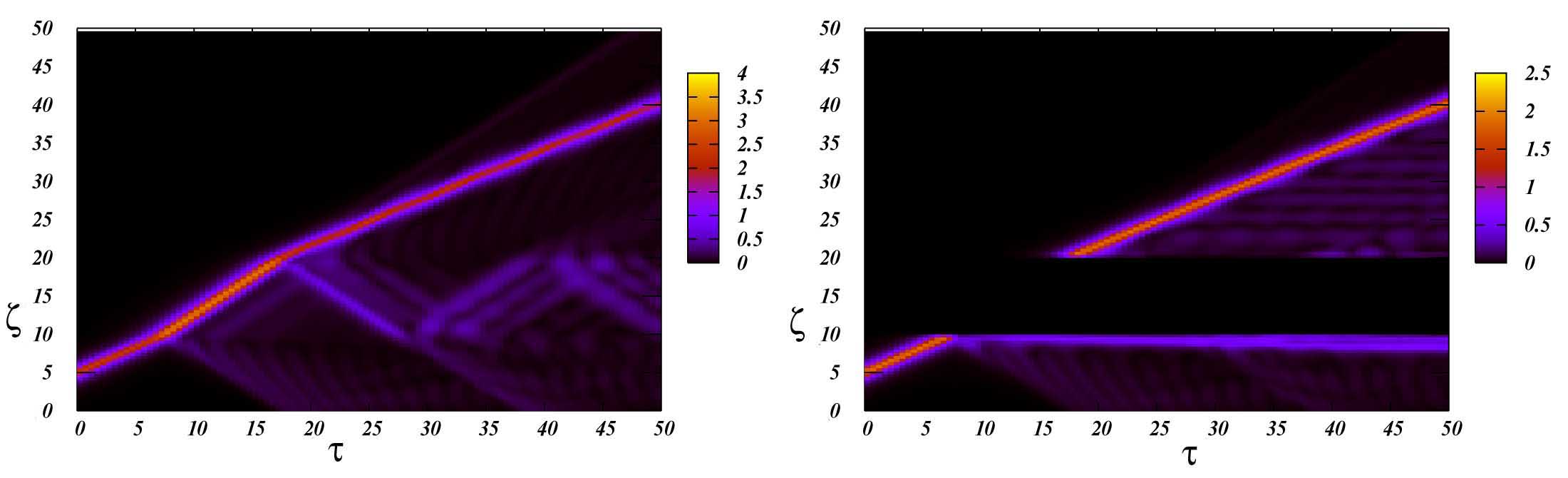}
\caption{{\protect\footnotesize {A typical example of PGS components
evolution in the RABG with microcavity. Total field is represented
on the left and polarization is on right. }}} \label{Fig2}
\end{figure}

In all following figures, which are illustrate results of numerical
simulations, the wide defect is represented by the span with
width=10 (normalized spatial units $\zeta$) with the indicator
function $\gamma (10<\zeta<20 )\neq 1$. Left part of these pictures
illustrates evolution of total electromagnetic field ($e_1+e_2$),
and right part is represents evolution of polarization component
$p$.

We consider the microcavity, i.e., wide defect which is not contains
resonant nanonarticles, as the first example of microresonator in
the RABG. At the front interface of the defect, as Fig.~\ref{Fig2}
demonstrates, the incident solitary polaritonic wave transforms to
the electromagnetic pulse which is propagates in the linear regime
until it crosses the second interface of the microcavity. (Inside
the defect the pulse velocity coincides with the velocity of linear
wave.) Near all energy of the incident electromagnetic pulse
transfers to this linear wave. Besides one can see the localized
plasmonic mode. It results from excitation of undamped plasmonic
oscillations due to polariton scattering on the microcavity edge.
Has traversed the cavity, the pulse sheds radiation on the second
edge of the defect. Since the reflected pulse has small
amplitude(less 0.2 - 0.4 of incident pulse amplitude), it transforms
into linear waves which are remain trapped in the cavity as they
cannot further propagate in RABG. The left part of Fig.~\ref{Fig2}
provides an illustration of linear radiation which is trapped by the
microcavity. The pulse refracted on the second interface restores
plasmonic wave, so nonlinear solitary polaritonic wave arises again
although with smaller energy.
\begin{figure}[t]
%\resizebox {width=8.5cm}{length=4.5cm}
\centering\includegraphics[width=8.5cm]{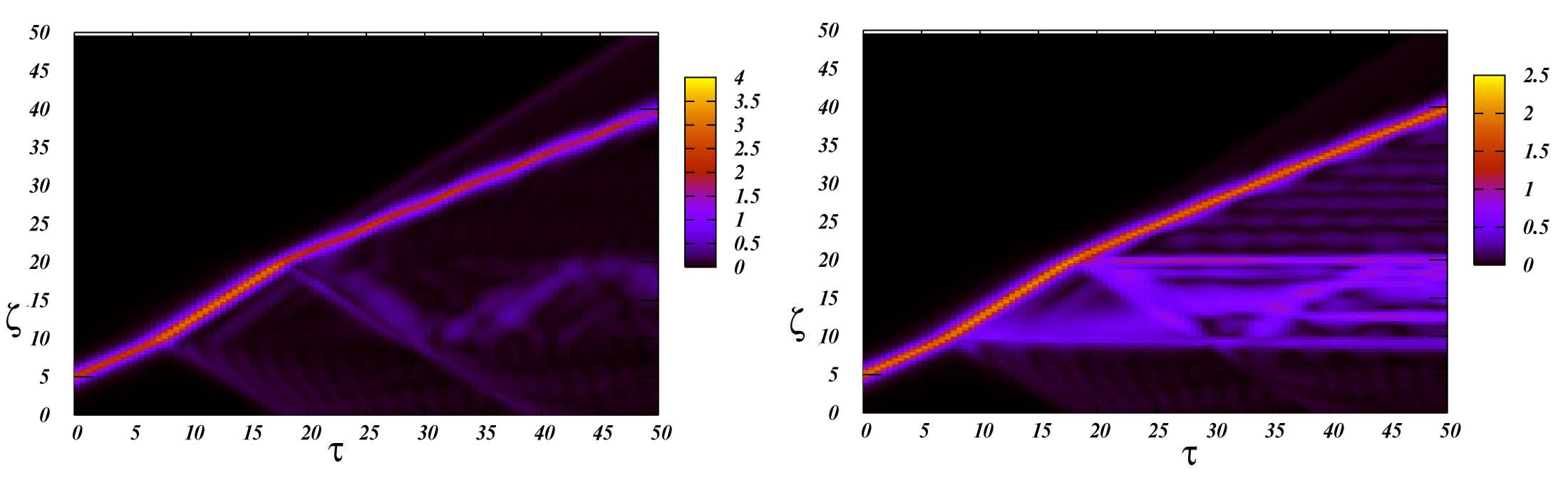}
\caption{{\protect\footnotesize {Polaritonic gap soliton ($\alpha
=1.5$) interaction with the wide groove with $\gamma=0.1$. }}}
\label{Fig3}
\end{figure}
Part of polaritonic pulse energy lost after passage the cavity is
lay out for generation the periodic localized polarization modes
that can be seen in the right of the Fig.~\ref{Fig2}. These bright
spots forming a periodic pattern are the consequence of reflection
between moving solitary pulse (moving mirror) and the edge of the
defect. We observed more pronounced periodic pattern for
polarization together with backward and forward waves (in total
field they are cancel each other due to interference) for the narrow
defect with $\gamma (10<\zeta<11 )=0$ (defect width is less then the
pulse width).

It is relevant to mention that the characteristic of the refracted
pulse passing through microcavity corresponds to light ray passing
the dielectric layer according to geometric optics laws.
%\begin{figure}[b]
% \centering\includegraphics[width=8.5cm]{Fig11.pdf}
%\caption{{\protect\footnotesize {Evolution of the total field (left)
%and the polarization (right) in the Bragg grating with a wide groove
%($\gamma (10<\zeta<20 )=0.25$).}}} \label{Fig11}
%\end{figure}
%\begin{figure}[b]
%\centering\includegraphics[width=8.5cm]{Fig12.pdf}
%\caption{{\protect\footnotesize {Evolution of the total field (left)
%and the polarization (right) in the Bragg grating with a wide groove
%($\gamma (10<\zeta<20 )=0.5$).}}} \label{Fig12}
%\end{figure}

Considering groove (i.e., region with decreased concentration of
resonant nanonarticles) one can see (Fig.~\ref{Fig3}) that part of
incident pulse energy is expended for excitation of the plasmonic
oscillations in the groove, hence the total electromagnetic field in
the cavity decreases (comparing to the microcavity case). Increase
the density of resonant nanonarticles ($\gamma=0.25$, $\gamma=0.5)$
in the span results in drop of the radiation losses inside the
groove-type defect. Linear polaritonic waves, which are trapped
inside the groove, remain conspicuous. In case then concentration of
nanoparticles in the groove is the half of the regular one,
polaritonic wave refraction at the fringes of the groove decreases,
and localization of plasmonic modes tends to occur close to the
fringes of the grove.

\begin{figure}[t]
%\resizebox {width=8.5cm}{length=4.5cm}
 \centering\includegraphics[width=8.5cm]{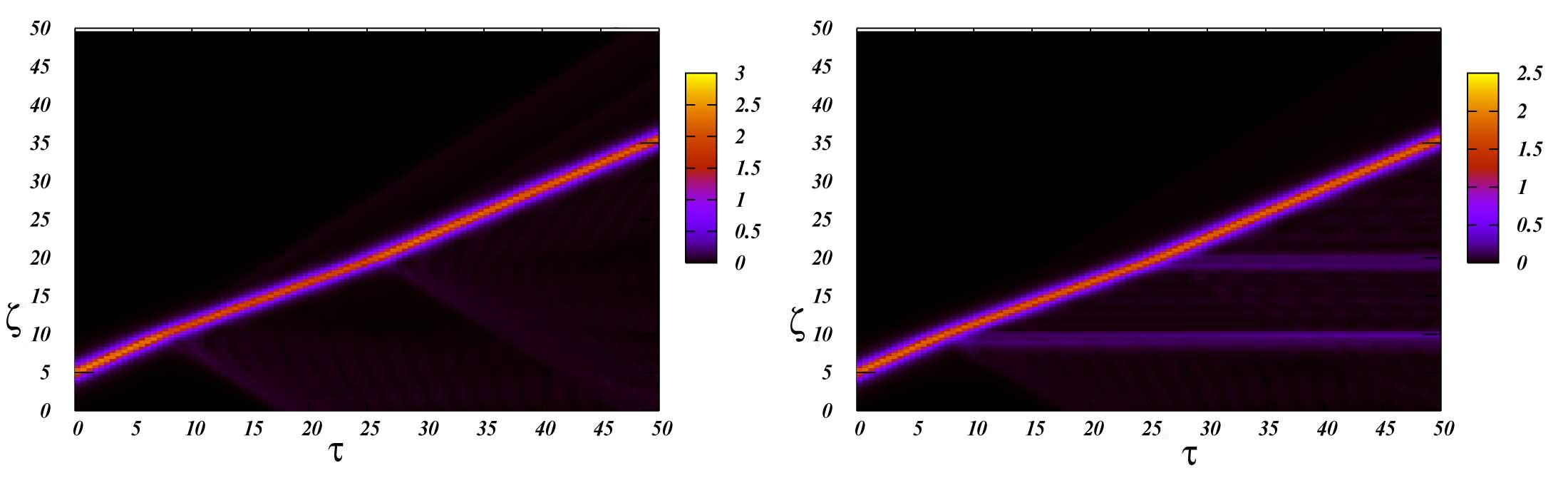}
\caption{{\protect\footnotesize {Polaritonic gap soliton ($\alpha
=1.5$) transmission through the wide stripe with $\gamma=1.5$. }}}
\label{Fig4}
\end{figure}

The third type of the microresonator is the stripe of optically
denser medium. Provided nanoparticle density in the stripe is not so
high ($\gamma=1.5$), solitary polaritonic wave undergoes small
losses of energy on the boundaries, and localization of polarization
modes occurs only on the fringes of the stripe (Fig.~\ref{Fig4}).
%\begin{figure}[b]
% \centering\includegraphics[width=8.5cm]{Fig13.pdf}
%\caption{{\protect\footnotesize {Evolution of the total field (left)
%and the polarization (right) in the Bragg grating with a wide stripe
%($\gamma (10<\zeta<20 )=3$).}}} \label{Fig13}
%\end{figure}
%\begin{figure}[b]
%\resizebox {width=8.5cm}{length=4.5cm}
% \centering\includegraphics[width=8.5cm]{Fig14.pdf}
%\caption{{\protect\footnotesize {Evolution of the total field (left)
%and the polarization (right) in the Bragg grating with a wide stripe
%($\gamma (10<\zeta<20 )=5$).}}} \label{Fig14}
%\end{figure}
\begin{figure}[b]
%\resizebox {width=8.5cm}{length=4.5cm}
 \centering\includegraphics[width=8.5cm]{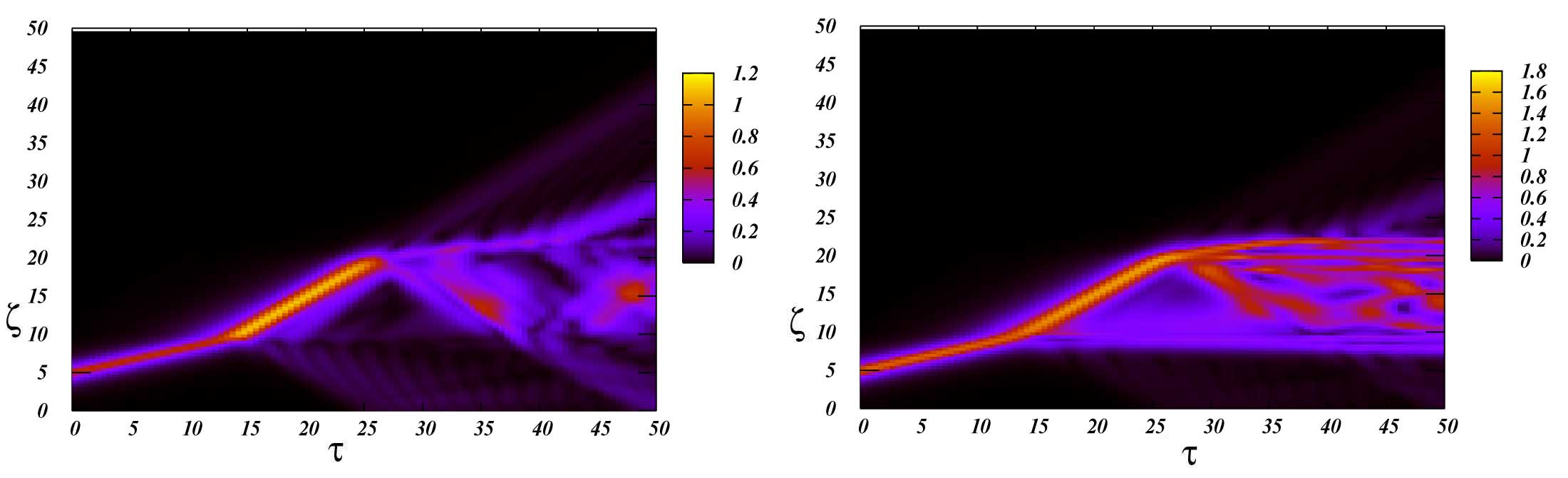}
\caption{{\protect\footnotesize {PGS ($\alpha=3$) interaction with
the wide groove($\gamma=0.1$) in the RABG.}}} \label{Fig5}
\end{figure}
\begin{figure}[b]
%\resizebox {width=8.5cm}{length=4.5cm}
 \centering\includegraphics[width=8.5cm]{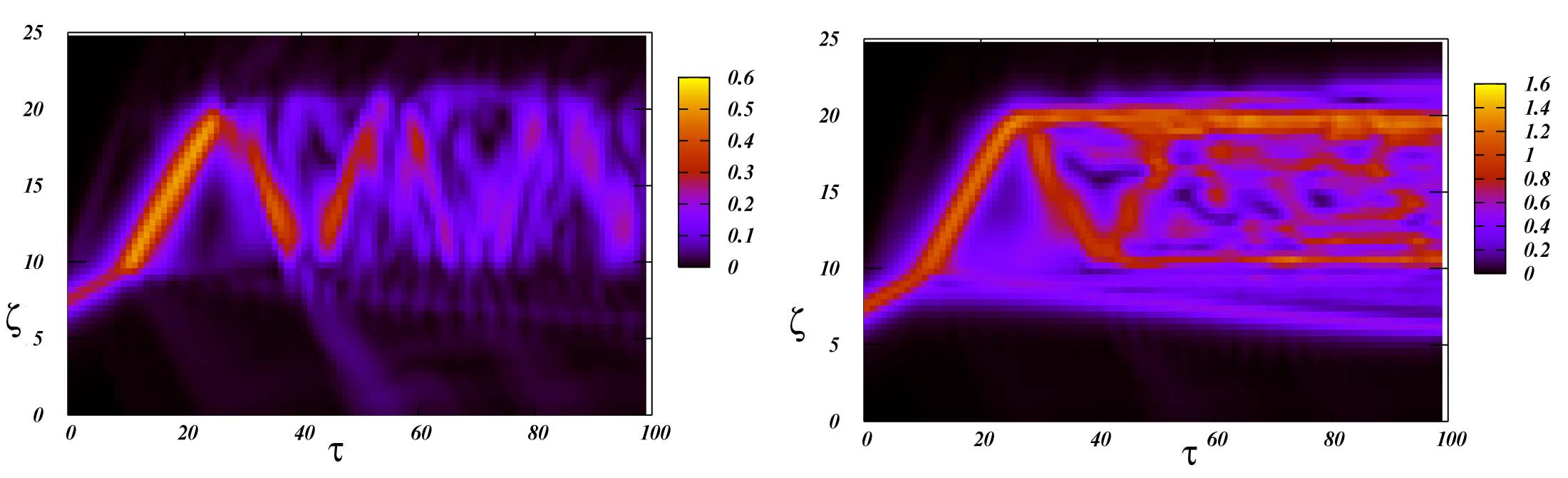}
\caption{{\protect\footnotesize {Slow PGS trapping in the
groove-type microresonator ($\gamma=0.1$).}}} \label{Fig6}
\end{figure}
At $\gamma=3$ the PGS effectively dissipates with appearance of
reflected backward waves. Some part of the incident pulse creates a
trapped plasmonic mode and the rest of pulse transfers forward being
converted into solitary wave with smaller amplitude. If density of
resonant nanoparticles in the defect is high, e.g., $\gamma=5$, the
PGS does not survive, most of its electromagnetic component reflects
backward. Also a high energetic plasmonic mode locking with the
fore-part of the stripe is appears. Part of radiation passes through
the stripe defect in the RABG. In the case of a narrow stripe of
dense material the transmitted pulse intensity is higher and the
pulse is more localized comparing with the pulse transmitted though
the wide stripe with the same density. Polarization mode is defined
only by the lower fringe reflection and it is seems does not change
with variation of the stripe width. As follows from (\ref{eq21}) and
definition of parameter $\eta$, solitary wave velocity is controlled
by parameter $\alpha$, i.e., increase of $\alpha$ leads to decrease
of PGS's velocity. Slow solitary wave is more effectively interact
with the defects. Systematic simulation have shown that in the case
of groove the slower PGS transfers more energy to localized
plasmonic oscillations (Fig.~\ref{Fig5}). If the polariton is very
slow, e.g., $\alpha=5$, it can be trapped effectively by the groove
(Fig.~\ref{Fig6}). Beyond the groove there is virtually no
radiation.

In the case of stripe almost all energy of the electromagnetic wave
transfers to the reflected pulse and near half energy of the
polaritonic wave transfers into the localized mode. Transmitted wave
is very weak.

In conclusion we have demonstrated that the wide defect in the RABG
can be considered as nonlinear microresonator. In particular weak
light pulse trapping is shown for the microcavity placed into RABG.

Solitary polaritonic wave scattering and transition is studied for
groove defects. We observed the capture of both electromagnetic
field and localized plasmonic mode in this case. Some part of
radiation remain localized in the groove. It is notable that the
slow PGS can be completely trapped by the groove-type
microresonator.

The near-total reflection resulted in PGS collapse on the
stripe-defect is demonstrated in numerical simulation. Generation of
localized plasmonic oscillations attends solitary wave scattering on
the defect. Stripe-type defect with high density of nanoparticles
could serve as the guide for highly intensive polarization mode
which propagates along the defect edge.

This work is funded by the State of Arizona grant TRIF-proposition
301 and in part by the Russian Foundation for Basic Research (grant
06-02-16406). We are grateful to Prof. Ildar Gabitov and Dr.
Alexander Korotkevich for enlightening discussions. Authors are
appreciate the hospitality and support of the Department of
Mathematics in the University of Arizona.

\end{document}